\documentclass{llncs}
\usepackage{makeidx}  % allows for indexgeneration
\usepackage{enumerate}
\usepackage{latexsym,makeidx,graphicx,color,colortbl}
\usepackage{moreverb,amsmath,amssymb}
\usepackage{pstricks,pst-node}
\usepackage[latin1]{inputenc}
\usepackage{tabulary}
\usepackage{multicol}
\usepackage{pdflscape}
\newcommand{\bdfn}{\begin{definition} \begin{rm}}
\newcommand{\edfn}{\vspace{-2ex}
{\flushright $\Box$\\
\mbox{}\vspace{-2ex}
} \end{rm} \end{definition}}
\newcommand{\bex}{\begin{example} \begin{rm}}
\newcommand{\eex}{\vspace{-2ex}{\flushright $\Box$\\
\mbox{}\vspace{-2ex}} \end{rm} \end{example}}

\newcommand{\entero}{\mathbb{Z}}

\newcommand{\nat}{{\rm I\! N}}

\newcommand{\reglaa}[3]{\renewcommand{\arraystretch}{0.5}
\begin{array}{c}
\displaystyle #1\\
\rule{#3cm}{0.1mm}\\
\displaystyle #2
\end{array}\renewcommand{\arraystretch}{1.0}
}

\newcommand{\flecha}[1]{\stackrel{#1}{\longrightarrow}}
\begin{document}
%
%\frontmatter          % for the preliminaries
%
%\pagestyle{headings}  % switches on printing of running heads
%\addtocmark{Hamiltonian Mechanics} % additional mark in the TOC
%
%\tableofcontents
%
\mainmatter              % start of the contributions
\title{BPEL-RF: A formal framework for BPEL orchestrations integrating distributed resources}

%\titlerunning{}  % abbreviated title (for running head)
\author{José Antonio Mateo \and Valent\'{\i}n Valero \and Gregorio D\'{\i}az}
\authorrunning{J. A. Mateo et al.}   % abbreviated author list (for running head)
%
%%%% modified list of authors for the TOC (add the affiliations)
%\tocauthor{Ivar Ekeland (Princeton University),
%Roger Temam (Universit\'{e} de Paris-Sud),
%Jeffrey Dean, David Grove, Craig Chambers (Universit\`a di Geova),
%Kim B. Bruce (Stanford University),
%Elisa Bertino (Digita Research Center)}
%
\institute{Informatics Research Institute of Albacete,\\
University of Castilla-La Mancha, Campus Universitario s/n,\\
02071. Albacete, SPAIN.\\
\email{\{valentin,jmateo,gregorio\}@dsi.uclm.es}}

\maketitle              % typeset the title of the contribution

\begin{abstract}
Web service compositions are gaining attention to develop complex web systems by combination of existing services. Thus, there are many works that leverage the advantages of this approach. However, there are only few works that use web service compositions to manage distributed resources. In this paper, we then present a formal model that combines orchestrations written in BPEL with distributed resources, by using WSRF.
\end{abstract}

\section{Introduction}
Software systems are gaining complexity and concurrency with the appearance
of new computational paradigms such as Service-Oriented  Computing  (SOC), Grid Computing and Cloud Computing.
In this kind of systems, the services provider needs to ensure some levels of quality and privacy to the final
user in a way that had never been raised. Therefore, it is necessary to develop new models yielding the advantages 
of recent approaches as web services compositions, but applied to these recent scenarios. To this end, we have worked up
an operational semantics to manage web services with associated resources by using the existing machinery in distributed systems, 
web services orchestrations.

The definition of a web service-oriented system involves 
two complementary views: Choreography and Orchestration. On the one hand, the choreography
concerns the observable interactions among services and can
be  defined  by  using  specific  languages,  e.g.,  Web  Services
Choreography  Description  Language  (WS-CDL) \cite{WSCDL}. On the other hand, the orchestration concerns
the internal behavior of a web service in terms of invocations
to  other  services.  Web Services Business Process Execution Language (WS-BPEL) \cite{BPEL4WS} is usually used 
to describe these orchestrations, so this
is considered the de facto standard language for describing web services
workflow in terms of web service compositions.

In this scenario, developers require more standardization to facilitate additional interoperability among these services. 
Thus, in January of 2004, several members of the organization \emph{Globus Alliance} and the computer multinational \emph{IBM}
with the help of experts from companies such as \emph{HP, SAP, Akamai, etc.} defined the basis architecture and the initial 
specification documents of a new standard for that purpose, Web Services Resource Framework (WSRF)~\cite{Foster2004}. Although the web service 
definition does not consider the notion of state, interfaces frequently provide the user with the ability to access and manipulate
states, i.e., data values that persist across, and evolve as a result of web service interactions. It is then desirable 
to define web service conventions to enable the discovery of, introspection on, and interaction with stateful resources in standard 
and interoperable ways~\cite{Czajkowski2004}. These observations motivated the appearance of the WS-Resource approach to modeling states in web services.

%A WS-Resource is defined as the composition of a Web service and a stateful resource that is (i) expressed as an association of an XML document with 
%defined type with a portType, and (ii) addressed and accessed according to the implied resource pattern, a conventional use
%of WS-Addressing endpoint references (EPR). In the implied resource pattern, a stateful resource identifier is encapsulated in an 
%endpoint reference and used to identify the stateful resource to be used in the execution of a Web service message 
%exchange. 
%An Endpoint Reference shall consist of: Uniform Resource Identifier (URI), parameters of the message being sent to request the sending of the Endpoint Reference and data on the interface used. In this exchange, the resource identifier is encapsulated in an Endpoint Reference used to identify the resource in any exchange of messages between services which belongs to the choreography. 

In WSRF, we can see a WS-Resource as a collection of properties \emph{P} identified by an address \emph{EPR} and with a \emph{timeout} associated. This timeout represents the lifetime of the WS-Resource. Without loss of generality, we have reduced the resource properties set to only one allowing us to use the resource identifier \emph{EPR} as the representative of this property. On the BPEL hand, we have only taken into consideration the root scope avoiding any class of nesting among scopes and we have only modeled the event and fault handling, leaving the other handling types as future work.

\section{Related Work}%\label{rWork}
The use of WS-BPEL has been extensively studied by using different types of formalism
such as Petri nets, Finite State Machines and process algebras. Regarding
the use of WS-BPEL together with WS-RF there are few works, and they only show a description of this union, 
without a formalization of the model.
In \cite{Slomiski:2006} Slomiski uses BPEL4WS in Grid environments and discusses the
benefits and challenges of extensibility in the particular case of OGSI workflows
combined with WSRF-based Grids. Other two works centered around Grid environments are 
\cite{Leymann:2006} and \cite{Ezenwoye:2007}. The first justifies 
the use of BPEL extensibility to allow the combination of different GRIDs, whereas
Ezenwoye et al.~\cite{Ezenwoye:2007} share their experience on BPEL to integrate,
create and manage WS-Resources that implement the factory/instance pattern.

On the Petri nets hand, Ouyang et al. \cite{Ouyang:2007} define the necessary elements for translating BPEL processes into Petri nets. Thus, they
cover all the important aspects in the standard such as exception handling, dead path elimination and so on. The model they consider differs from ours in that we formalize the whole
system as a composition of orchestrators with resources associated, whereas they describe the system as a general scope with nested sub-scopes leaving aside the possibility of administering resources. Furthermore, we have also formalized the event handling and notification mechanisms.  
Another extensive semantics for BPEL 2.0
is presented in \cite{Dumas:2008} by Dumas et al, which introduces two new interesting improvements. They define several patterns to simplify some huge nets and introduce the semantics for the WS-BPEL 2.0 new patterns. On the $\pi$-calculus hand, Dragoni and Mazzara \cite{Dragoni:2009} 
propose a theoretical scheme focused on dependable composition for the  WS-BPEL recovery
framework. In this approach, the recovery framework is simplified
and analyzed via a conservative extension of $\pi$-calculus. The
aim of this approach clearly differs from ours, but it helps us to have
a bigger understanding of the WS-BPEL recovery framework. Other work focused on the BPEL
recovery framework is \cite{Qiu:2005}. Although this is more interested in the compensation handler, they describe the corresponding rules
that manage a web service composition. Our work is therefore quite complete as we define rules for nearly all possible activities. In addition, we also consider time constraints. Finally, we would like to highlight the works of Farahbod et al.~\cite{Farahbod:2005} and Busi et al. \cite{Busi:2005}. In the first one, the authors extract an abstract operational semantics for BPEL based on abstract state machines (ASM) defining the framework BPEL$_{AM}$ to manage the agents who perform the workflow activities. In this approach time constraints are considered, but they do not formalize the timed model. On the other hand, the goal of the latter one is fairly similar to ours. They also define a $\pi$-calculus operational semantics for BPEL and describe a conformance notion. They present all the machinery to model web service compositions (choreographies and orchestrations). The main differences with our work are that we are more restrictive with respect to time constraints and we deal with distributed resources.

\section{BPEL/WSRF}\label{BPEL/WSRF}

WS-Resource Framework~\cite{Banks2006} is a resource specification language developed by OASIS and some of the most pioneering computer companies, whose purpose is to define a generic framework for modeling web services with stateful resources, as well as the relationships among these services in a Grid/Cloud environment. This approach consists of a set of specifications that define the representation of the WS-Resource in the terms that specify the messages exchanged and the related XML documents. These specifications allow the programmer to declare and implement the association between a service and one or more resources. It also includes mechanisms to describe the means to check the status and the service description of a resource, which together form the definition of a WS-Resource. In Table~\ref{tabla} we show the main WSRF elements. %These conventional mechanisms are described as follows:
%The most important thing is to know how this approach improves robustness in the selection of services during the assembly of the application and how to connect those services to the instances of their associated resources. These observations have led to the appearance of a proposal discussed above, WS-Resource. As mentioned earlier,  In this sense, WSRF is useful to declare, create, access, monitoring and destroying WS-Resources through conventional mechanisms, which makes it easy to manage, since it is not necessary to make more difficult the decision logic of the service owner of the resource to manage the messages exchanges.

%\begin{figure}[!h]
%\begin{center}
%\includegraphics[scale=0.25]{Figures/tabla.eps}
%\end{center}
%\vspace{-0.4cm}
%\caption{WSRF specifications.} \label{tabla}
%\end{figure}
%\vspace{-0.5cm}
{\renewcommand{\arraystretch}{0.85}
\begin{table}[!h]
{
\scriptsize

\begin{center}
\begin{tabular}{|p{3.5cm}|p{8.5cm}|}
\hline
\cellcolor[gray]{.9}~ & \cellcolor[gray]{.9}~ \\
%\rowcolor[gray]{.9}
%\multicolumn{1}{>{\rowcolor[gray]{.9}}c|}{WS-CDL Syntax} & Metamodel\\
\cellcolor[gray]{.9}\hspace*{1.5cm}Name & \cellcolor[gray]{.9}\hspace*{3.5cm}Describes\\
\cellcolor[gray]{.9}~ & \cellcolor[gray]{.9}~ \\
\hline

\begin{center}
\vspace{-0.3cm}
{\bf WS-ResourceProperties} 
\end{center}
& 
%\begin{flushleft}
\vspace{0.0cm}
\hspace{0.3cm} WSRF uses a precise specification to define the properties of the WS-Resources.\\
%\end{flushleft}\\
\hline

\begin{center}
\vspace{-0.35cm}
{\bf WS-Basefaults} 
\end{center}
& 
%\begin{flushleft}
\vspace{0.0cm}
\hspace{0.3cm} To standardize the format for reporting error messages.
\\[-0.2cm]
%\end{flushleft}\\
\hline

\begin{center}
\vspace{-0.3cm}
{\bf WS-ServiceGroup} 
\end{center}
& 
\begin{flushleft}
\vspace{-0.4cm}
\hspace{0.3cm} This specification allows the programmer to create groups that share a common set of properties.
\end{flushleft}\\[-0.2cm]
\hline

\begin{center}
\vspace{-0.15cm}
{\bf WS-ResourceLifetime}
\end{center}
& 
\begin{flushleft}
\vspace{-0.4cm}
\hspace{0.3cm} The mission of this specification is to standardize the process of destroying a resource and identify mechanisms 
to monitor its lifetime. 
\end{flushleft}\\[-0.2cm]
\hline

\begin{center}
%\vspace{-0.1cm}
{\bf WS-Notification}
\end{center}
& 
\begin{flushleft}
\vspace{-0.3cm}
\hspace{0.3cm} This specification allows to a \emph{NotificationProducer} to send \emph{notifications} 
to a \emph{NotificationConsumer} in two ways: without following any formalism or with a predefined formalism.
\end{flushleft}\\%[-0.1cm]
\hline

\end{tabular}
\end{center}
}
%}% Fin de fbox
\caption{\label{tabla}WSRF main elements}
\end{table}}

On the other hand, web services are becoming more and more important as a platform
for Business-to-Business integration.  Web service compositions have appeared
as a natural and elegant way to provide new value-added services
as a combination of several established web services.
Services provided by different suppliers can act together
to provide another service; in fact, they can be written in different
languages and can be executed on different platforms. As we noticed in the introduction, we can use web service compositions as a way to construct web service systems where each service is an autonomous entity which can offer a series of operations to the other services conforming a whole system. In this way, it is fairly necessary to establish a consistent manner to coordinate the system participants such that each of them may have a different approach, so it is common to use specific languages such as WS-BPEL to manage the system workflow. WS-BPEL, for short BPEL, is an OASIS orchestration language for specifying actions within web service business processes. These actions are represented by the execution of two types of activities (\emph{basic} and \emph{structured}) that perform the process logic. \emph{Basic activities} are those which describe elemental steps of the process behavior and \emph{structured activities} encode control-flow logic, and can therefore contain other basic and/or structured activities recursively~\cite{BPEL4WS}. %Of those present in the first group are: \emph{invoke} to call services offered by service providers, \emph{receive} and \emph{reply} to provide services to partners, \emph{throw} to signal an internal fault explicitly, \emph{wait} to specify a delay for a certain period of time, \emph{empty} to do nothing, \emph{exit} to end the business process and \emph{assign}, which is used to copy data from a variable to another. On the other hand, among the \emph{structured activities} are: \emph{sequence} which contains one or more activities that are performed sequentially, \emph{if} to provide a conditional behavior, \emph{while} to provide repeated execution of a contained activity,  \emph{repeatuntil} with a similar behavior to the previous one, \emph{pick} that waits for the occurrence of exactly one event from a set of events (including alarm event), and then executes the activity associated with that event, \emph{flow} to express concurrency and synchronization, and, finally, \emph{foreach} to execute the same activity a predetermined number of times.

\section{Operational Semantics}

We use the following notation: {\it ORCH} is the set of orchestrators in the system, {\it Var} is the set of integer variable names, {\it PL} is the set of necessary partnerlinks, {\it OPS} is the set of operations that can be performed, {\it EPRS} is the set of resource identifiers, and {\it A} is the set of basic or structured activities that can form the body of a process. The specific algebraic language, then, that we use
for the activities is defined by the following BNF-notation:
\[\begin{array}{l}
  A ::=  {\it throw} \;|\;           % go to the exception block
         {\it receive}(pl,op,v) \;|\;  % receive basic activity,
         {\it invoke}(pl,op,v_1) \;|\;\\
         %~~~
         {\it reply}(pl,v)\;|\;  % reply basic activity
         {\it \overline{reply}(pl,v_2)}\;|\; 
         {\it assign}(expr,v_1) \;|\;
         {\it wait}(timeout)\hspace{-0.1cm}\;|\;\\
         %~~~~~~~
         {\it empty}\;|\;
         {\it exit}\;|\;\
         \,\,A \,; A \,\, \;|\, % Sequence
         \;A \,\| \, A \;\,|\, 
            % parallel
         {\it while}(cond,A)\;|\\\
         {\it pick}(\{(pl_i,op_i,v_i,A_i)\}_{i=1}^n,A,timeout)\;|\;\\
         {\it createResource}(EPR,val,timeout,A_{e{_i}})\;|\;
         {\it getProp}(EPR,v)\hspace{-0.1cm}\;|\;\\
         {\it setProp}(EPR,val)\;|\;
         {\it setTimeout}(EPR,timeout)\;|\;\\
         {\it subscribe}(O,EPR,cond',A_{e{_i}})
\end{array}
\]
where ${\it O \in ORCH, EPR \in EPRS , pl,pl_{i} \in PL, op,op_{i}}$ ${\it \in OPS, timeout \in \nat, expr}$ is an arithmetic expression constructed by using the variables in {\it Var} and integers; ${\it v,v_1,v_2,v_i}$ range over {\it Var}, and ${\it val \in \entero}$. A condition ${\it cond}$ is a predicate constructed by using conjunctions, disjunctions, and negations over
the set of variables ${\it Var}$ and integers, whereas ${\it cond'}$ is a predicate constructed by using the corresponding ${\it EPR}$ (as the resource value) and integers.  

BPEL basic activities used in our model are: \emph{invoke} to request services offered by service providers, \emph{receive} and \emph{reply} to provide services to partners, \emph{throw} to signal an internal fault explicitly, \emph{wait} to specify a delay, \emph{empty} to do nothing, \emph{exit} to end the business process and \emph{assign}, which is used to copy data from a variable to another. And the \emph{structured activities} used are: \emph{sequence}, which contains two activities that are performed sequentially, \emph{while} to provide a repeated execution of one activity, \emph{pick} that waits for the occurrence of exactly one event from a set of events (including an alarm event), and then executes the activity associated with that event, and, finally, \emph{flow} to express concurrency. Another family of control flow constructs in BPEL includes event, fault and compensation handlers. An event handler is enabled when
its associated event occurs, being executed concurrently with the main orchestrator activity. Unlike event handlers, fault handlers do not execute concurrently with the orchestrator main activity~\cite{Ouyang:2007}. The correspondence among the syntax of WS-BPEL, WSRF and our model is shown in Table \ref{BPELsyntax}.

%\newpage

{\renewcommand{\arraystretch}{0.85}
\begin{table}[!h]
{
\tiny

\begin{center}
\begin{tabular}{|p{8.5cm}|p{4.5cm}|}
\hline
\cellcolor[gray]{.9}~ & \cellcolor[gray]{.9}~ \\
%\rowcolor[gray]{.9}
%\multicolumn{1}{>{\rowcolor[gray]{.9}}c|}{WS-CDL Syntax} & Metamodel\\
\cellcolor[gray]{.9}\hspace*{3.3cm}WS-BPEL Syntax & \cellcolor[gray]{.9}\hspace*{1.7cm}Metamodel\\
\cellcolor[gray]{.9}~ & \cellcolor[gray]{.9}~ \\
\hline

\begin{flushleft}
\vspace{-0.2cm}
$<$process ...$>$\\
~~$<$partnerLinks$>$ ... $<$/partnerLinks$>$?\\
~~$<$Variables$>$ ... $<$/Variables$>$?\\
~~$<$faultHandlers$>$ ... $<$/faultHandlers$>$?\\
~~$<$eventHandlers$>$ ... $<$/eventHandlers$>$?\\
~~~~~(activities)*\\
$<$/process$>$\\
%~
\end{flushleft}
& 
\begin{center}
\vspace{0.2cm}
(PL,Var,A,A$_f$,$\mathcal{A}_e$)
\end{center}\\
\hline

\begin{tabular}{l}
~\\
throw/any fault
\end{tabular}
& \begin{center}
\vspace{-0.4cm}
throw
\end{center}\\
\hline

\begin{tabular}{l}
~\\
$<$receive partnerLink=``pl'' 
operation=``op''
variable=``v''
createInstance=``no''$>$\\
$<$/receive$>$\\
~\end{tabular}
& 
\begin{center}
\vspace{-0.5cm}
receive(pl,op,v)
\end{center}\\
\hline

\begin{tabular}{l}
~\\
$<$reply partnerLink=``pl'' variable=``v''$>$
$<$/reply$>$\\
~\end{tabular}
& 
\begin{center}
\vspace{-0.5cm}
reply(pl,v)
\end{center}\\
\hline

\begin{tabular}{l}
~\\
$<$invoke partnerLink=``pl'' operation=``op''inputVariable=``v$_{1}$''\\
outputVariable=``v$_{2}$''$>$
$<$/invoke$>$\\
~\end{tabular}
&

\begin{center}
\vspace{-0.5cm}
invoke(pl,op,v$_{1}$);
$[\overline{reply}$(pl,op,v$_2$)]\\

\end{center}\\
\hline

\begin{tabular}{l}
~\\
$<$empty$>$~\ldots~$<$/empty$>$\\
~\end{tabular}
& \begin{center}\vspace{-0.5cm}empty\end{center}\\
\hline

\begin{tabular}{l}
~\\
$<$exit$>$~\ldots~$<$/exit$>$\\
~\end{tabular}
& \begin{center}\vspace{-0.5cm}exit \end{center}\\
\hline

\begin{tabular}{l}
~\\
$<$assign$>$$<$copy$>$$<$from$>$expr$<$/from$>$$<$to$>$v$_1$$<$/to$>$$<$/copy$>$$<$/assign$>$\\
~\\
\end{tabular}
& 
\begin{center}
\vspace{-0.5cm}
assign(expr,v$_{1}$)
\end{center}\\
\hline

\begin{tabular}{l}
~\\
$<$wait$>$$<$for$>$timeout$<$/for$>$ $<$/wait$>$\\
~\end{tabular}
& 
\begin{center}
\vspace{-0.5cm}
wait(timeout)
\end{center}\\
\hline

% SEQUENCE AND PARALLEL
%
\begin{tabular}{c|c}
%
% SEQUENCE
\begin{tabular}{l}
~\\
$<$sequence$>$\\
~~~activity$_1$\\
~~~activity$_2$\\
$<$/sequence$>$\\
~
\end{tabular}
&
% PARALLEL
%
~~\begin{tabular}{l}
~\\
$<$flow$>$\\
~~~activity$_1$\\
~~~activity$_2$\\
$<$/flow$>$\\
~
\end{tabular}~~
\end{tabular}
&
\begin{tabular}{c}
~\\
\hspace{1.5cm}A$_1 \,; \,$ A$_2$\\
~\\
\hspace{1.5cm}-----------------
~\\
\hspace{1.5cm}A$_1 \,\| \,$ A$_2$\\
~\\
\end{tabular}\\
\hline

\begin{tabular}{l}
~\\
$<$while$>$$<$condition$>$cond$<$/condition$>$activity$_1$$<$/while$>$\\
~\end{tabular}
& \begin{center}\vspace{-0.5cm}while(cond,A)\end{center}\\
\hline

\begin{tabular}{l}
~\\
$<$pick createInstance=``no''$>$\\
~~$<$onMessage partnerLink=``pl'' operation=``op''variable=``v''$>$\\
~~~~activity$_1$\\
~~$<$/onMessage$>$\\
~~$<$onAlarm$>$$<$for$>$timeout$<$/for$>$activity$_1$$<$/onAlarm$>$ \\
$<$/pick$>$\\
~\end{tabular}
& \begin{center}\vspace{-0.5cm}pick($\{(pl_i,op_i,v_i,A_i)\}_{i=1}^n,A$,timeout)\end{center}\\
\hline

\begin{tabular}{l}
~\\
$<$invoke partnerLink=``Factory''operation=``CreateResource''\\
inputVariable=``MessageIn''outputVariable=``MessageOut''$>$\\
$<$/invoke$>$$<$assign$>$$<$copy$>$$<$from variable=``MessageOut''$>$part=``param''\\
query=``/test:CreateOut/wsa:endpointreference''$<$/from$>$\\ 
$<$to$>$ partnerlink=``Factory''$<$/to$>$$<$/copy$>$$<$/assign$>$\\
~\end{tabular}
& \begin{center}\vspace{-0.5cm}createResource(EPR,val,timeout,A$_{e{_i}}$)\end{center}\\
\hline

\begin{tabular}{l}
~\\
$<$wsrp:GetResourceProperty$>$property$_1$$<$/wsrp:GetResourceProperty$>$\\
\end{tabular}
& \begin{center}\vspace{-0.48cm}getProp(EPR,v)\end{center}\\
\hline

\begin{tabular}{l}
~\\
$<$wsrp:SetResourceProperties$>$\\
~~$<$wsrp:Update$>$~~property$_1$~~$<$/wsrp:Update$>$ \\
$</$wsrp:SetResourceProperties$>$\\
~
\end{tabular}
&
\begin{center}\vspace{-0.5cm}setProp(EPR,val)\end{center}\\
\hline

\begin{tabular}{l}
~\\
$<$wsrl:SetTerminationTime$>$\\
~~$<$wsrl:RequestedTerminationTime$>$\\
~~~~timeout\\
~~$<$/wsrl:RequestedTerminationTime$>$\\
$<$/wsrl:SetTerminationTime$>$\\
~
\end{tabular}
& \begin{center}\vspace{-0.5cm}setTimeout(EPR,timeout)\end{center}\\
\hline

\begin{tabular}{l}
~\\
$<$wsnt:Subscribe$>$\\
~~$<$wsnt:ConsumerReference$>$O$<$/wsnt: ConsumerReference$>$\\
~~$<$wsnt:ProducerReference$>$EPR$<$/wsnt: ProducerReference$>$\\
~~$<$wsnt:Precondition$>$cond'$<$/Precondition$>$\\
$<$/wsnt:Subscribe$>$\\
~
\end{tabular}
& \begin{center}\vspace{-0.5cm}subscribe(O,EPR,cond',A$_{e{_i}}$)\end{center}\\
\hline

% Notify esta dentro del metamodelo?¿?

\begin{tabular}{l}
~\\
$<$wsnt:Notify$>$\\
~$<$wsnt:NotificationMessage$>$\\
~~$<$wsnt:SubscriptionReference$>$O$<$/wsnt:SubscriptionReference$>$\\ 
~~$<$wsnt:ProducerReference$>$$EPR$$<$/wsnt:ProducerReference$>$\\ 
~~$<$wsnt:Message$>$~~...~~$<$/wsnt:Message$>$\\
~$<$/wsnt:NotificationMessage$>$\\
$<$/wsnt:Notify$>$\\
~
\end{tabular}
& \hspace{0.2cm}Spawn the associated event handler activity A$_{e{_i}}$\\
\hline

\end{tabular}
\end{center}
}
%}% Fin de fbox
\caption{\label{BPELsyntax}Conversion table}
\end{table}}
%\newpage

An orchestration is now defined as a tuple ${\it O= (PL,Var,A,A_f,\mathcal{A}_e)}$, where $A$ and $A_f$ are activities defined by the previous syntax and $\mathcal{A}_e$ is a set of activities. Specifically, $A$ represents the normal workflow, $A_f$ is the fault handling activity and $\mathcal{A}_e=\{A_{e_{i}}\}_{i=0}^m$ are the event handling activities. The operational semantics is, then, defined at three levels, the internal one corresponds to the evolution of one activity without notifications. In the second one, we define the orchestration semantics with notifications, whereas the third level corresponds to the composition
of different orchestrators and resources to conform the choreography. We first introduce some definitions that are required in order to define the operational semantics.

\begin{definition}[States]\begin{rm}
\label{states}
We define a state as a pair s=($\sigma, \rho$), where $\sigma$ represents the variable values and $\rho$ captures the resource state. Thus, \mbox{${\it \sigma:Var \rightarrow \entero}$}, and \linebreak $\it{ \rho=\{(EPR_i,v_i,Subs_i,t_i, A_{e{_i}})\}_{i=1}^r}$, where $r$ is the number of resources in the system. Each resource has its own identifier, ${\it EPR_i}$, and, at each state, has a particular value, $v_i$, and a lifetime, $t_i$, initialized with the {\it createResource} function, which can be changed by using the function {\it setTimeout}. Moreover, $\it{Subs_i=\{(O_{i_{j}},cond'_{i_{j}},A_{e{_s{_{_i{_{_j}}}}}})\}_{j=1}^{s_{i}}}$ is the set of resource notification subscribers, their associated delivery conditions and the event handling activity ${\it A_{e{_s{_{_i{_{_j}}}}}}}$ that must be thrown in the case that ${\it cond'_{i_{j}}}$ holds; $s_i$ is the number of orchestrations currently subscribed to this resource and ${\it O_{i_{j}} \in ORCH}$ are the subscriber's identifiers. The operations are defined as follows: ${\it OPS=\{op_i|\ op_i:\entero^{Var}\rightarrow\entero^{Var}\}}$. Given a state $s=(\sigma,\rho)$, a variable $v$ and an expression $e$,
we denote by $s'=(\sigma[e/v],\rho)$ the state obtained from $s$
by changing the value of $v$ for the evaluation of $e$ and ${\it s{^+}=(\sigma,\rho')}$, where ${\it \rho'=\{(EPR_i,v_i,Subs_i,t_i-1,A_{e{_i}}) | t_i>1\}_{i=1}^r}$. 
%\vspace{-0.2cm}
\edfn
%\end{definition}
\vspace{-0.3cm}
Next we define some notation that we use in the operational semantics. We employ the notation $\it{EPR_i \in \rho}$ to denote that there is a tuple $\it{(EPR_i,v_i,Subs_i,t_i,A_{e{_i}})}$ \\${\it \in \rho}$, $i \in [1\ldots r]$. Given a predicate $\it{cond}$, we use the function $\it{cond(s)}$ to mean the resulting value of this predicate at the state $\it{s}$. Besides, ${\it \rho[w/EPR]_1}$ is used to denote that the new value in $\rho$ of the resource $\it{EPR}$ is $\it{w}$, $\it{\rho[t/EPR]_2}$ denotes a change in the $\it{timeout}$ attribute of the resource in $\rho$ and \\$\it{Add\_subs(\rho, EPR_i,O_{i_{j}},cond'_{i_{j}},A_{e{_s{_{_i{_{_j}}}}}})}$ denotes that $\it{(O_{i_{j}},cond'_{i_{j}},A_{e{_s{_{_i{_{_j}}}}}})}$ is added to the subscribers of the resource $\it{EPR_i \in \rho}$ or ${\it cond'=cond'_{i_{j}}}$ in the case that $\it{O_{i_{j}}}$ was already in ${\it Subs_i}$. We need two additional functions. One of them, to extract the event handling activities that will be launched 
%in parallel with the normal workflow 
when the subscriber condition holds at the current state ${\it s}$: 
${\it N(O,s)=\{A_{e{_s{_{_i{_{_j}}}}}}|(O_{i_{j}},cond'_{i_{j}},A_{e{_s{_{_i{_{_j}}}}}}) \in Subs_i, O_{i{_j}}=O,}$\\ ${\it cond'_{i_{j}}=true\}_{i=1}^r}$ and the other one is used to launch the activities when the resource lifetime expires: ${\it T(O,s)=\{ A_{e{_r{_{_i}}}}|(EPR_i,v_i,Subs_i,1,A_{e{_r{_{_i}}}}) \in \rho, O=}$\\ ${\it O_{i{_j}}\in Subs_i\}_{i=1}^r}$. Now, a partnerlink is a pair $(O_i,O_j)$ representing the two roles in communication: sender and receiver.
\vspace{0.1cm}   
\begin{definition}[Activity Operational semantics]\begin{rm}
We specify the activity operational semantics by using two types of transition:

\begin{enumerate}
\item \hspace{0.1cm}(A,s)$\xrightarrow{a}(A',s')$, a $\in$ Act \hspace{0.3cm}(Action transitions).
\item \hspace{0.1cm}(A,s)$\xrightarrow{}_1(A',s^+)$ \hspace{1.4cm}(Delay transitions).
\end{enumerate}

\edfn
\vspace{-0.3cm}
\noindent where Act is the set of actions that can be performed, namely:  $Act=\{\tau$, throw, receive(pl,op,v), reply(pl,v), invoke(pl,op,v$_1$),
$\overline{reply}$(pl,v$_2$), assign(e,v$_1$), empty,\\ wait(timeout), exit, pick(\{(pl$_i$,op$_i$,v$_i$,A$_i$)\}$_{i=1}^n$,A,timeout), while(cond,A), \\ createResource(EPR,val,timeout,A$_{e_{i}}$), setProp(EPR,val), getProp(EPR,v), setTimeout(EPR,timeout),
and subscribe(O,EPR,cond$'$,A$_{e{_i}})$\}. Notice that we have included a ${\it \tau}$-action that represents an empty movement.

\emph{Action transitions} capture a state change by the execution of an action $a \in Act$, which can be empty ($\tau$). \emph{Delay transitions}
capture how the system state changes when one time unit has elapsed. In Tables \ref{tran1},\ref{tran2}, we show the rules of these transitions.

\begin{table}[!ht]
{\scriptsize
\framebox[12.3cm]{
\begin{tabular}{l}

\hspace{1cm}\textbf{(Throw)}
${ % Consecuente
(throw,s)
\xrightarrow{throw}
(empty, s)
}$

\hspace{0.5cm}
\textbf{(Exit)} 
$ % Consecuente
(exit,s)
\xrightarrow{exit}
(empty,s)
$
\\

\hspace{1cm}\textbf{(Receive)}
${ % Consecuente
(receive(pl,op,v),s)
\xrightarrow{receive(pl,op,v')}
(empty, s')
}$\\
\hspace{1cm}where  ${\it v \in Var, v'\in \entero, op \in OPS, pl \in PL}$, and ${\it s'=(op(\sigma[v'/v]),\rho)}$.

\\%[0.2cm]

\hspace{0.9cm}
\textbf{(Invoke)}
$
{ % Consecuente
(invoke(pl,op,v_1),s)
\xrightarrow{invoke(pl,op,v_1)}
(empty,s)
}$
%where ${\it v_1, v_2 \in Var, op \in OPS}$, and $\it{pl \in PL}$.
\\%[0.2cm]

\hspace{1cm}${\bf (\overline{Reply})}$
$
{ % Consecuente
(\overline{reply}(pl,v_2),s)
\xrightarrow{\overline{reply}(pl,v'_2)}
(empty,s')
}$\\
\hspace{1cm}where  ${\it v_2 \in Var, v'_2 \in \entero, pl \in PL}$, and ${\it s'=(\sigma[v'_2/v_2],\rho)}$. 
\\

\hspace{1cm}\textbf{(Reply)}
$
{ % Consecuente
(reply(pl,v), s)
\xrightarrow{reply(pl,v)}
(empty,s)
}$
\\

\hspace{1cm}\textbf{(Assign)}
$
{ % Consecuente
(assign(expr,v_1),s)
\xrightarrow{assign(expr,v_1)}
(empty,s')
}$\\
\hspace{1cm}where ${\it v_1 \in Var, expr}$ is an arithmetic expression, and ${\it s'=(\sigma[expr/v_1],\rho)}$.
\\

\hspace{1cm}\textbf{(Seq1)}
$\reglaa{ % Antecedente
(A_1,s)\xrightarrow{a}(A'_1,s'), a\neq exit, a\neq throw         
}
{ % Consecuente
(A_1;A_2,s)
\xrightarrow{a}
(A'_1;A_2,s')
}{5}$

\\

\hspace{1cm}\textbf{(Seq2)}
$\reglaa{ % Antecedente
(A_1,s)\xrightarrow{a}(empty,s'), a\neq exit,  a\neq throw 
}
{ % Consecuente
(A_1;A_2,s)
\xrightarrow{a}
(A_2,s')
}{5}$

%where ${\it s'=(\sigma',\rho)}$ or ${\it s'=(\sigma,\rho')}$.\\
\\

\hspace{1cm}\textbf{(Seq3)}
$\reglaa{ % Antecedente
(A_1,s)\xrightarrow{a}(empty,s), (a=throw \vee a=exit)    
}
{ % Consecuente
(A_1;A_2,s)
\xrightarrow{a}
(empty,s)
}{5}$
\\

\hspace{1cm}\textbf{(Par1)}
$\reglaa{ % Antecedente
(A_1,s)\xrightarrow{a}(A'_1,s'), a\neq exit, a\neq throw             
}
{ % Consecuente
(A_1||A_2,s)
\xrightarrow{a}
(A'_1||A_2,s')
}{5}$

%where ${\it s'=(\sigma',\rho)}$ or ${\it s'=(\sigma,\rho')}$.\\

\\

\hspace{1cm}\textbf{(Par2)}
$\reglaa{ % Antecedente
(A_2,s)\xrightarrow{a}(A'_2,s'), a\neq exit, a\neq throw            
}
{ % Consecuente
(A_1||A_2,s)
\xrightarrow{a}
(A_1||A'_2,s')
}{5}$

%where ${\it s'=(\sigma',\rho)}$ or ${\it s'=(\sigma,\rho')}$.\\

\\

\hspace{1cm}\textbf{(Par3)}
$\reglaa{ % Antecedente
(A_1,s)\xrightarrow{a}(empty,s), (a=throw \vee a=exit)            
}
{ % Consecuente
(A_1||A_2,s)
\xrightarrow{a}
(empty,s)
}{5}$

\\

\hspace{1cm}\textbf{(Par4)}
$\reglaa{ % Antecedente
(A_2,s)\xrightarrow{a}(empty,s), (a=throw \vee a=exit)                
}
{ % Consecuente
(A_1||A_2,s)
\xrightarrow{a}
(empty,s)
}{5}$

\\
\hspace{1cm}\textbf{(Par5)}
${ % Consecuente
(empty||empty,s)
\xrightarrow{\tau}
(empty,s)
}$
\\

\hspace{1cm}\textbf{(While1)}
$\reglaa{ % Antecedente
cond(s)  
}
{ % Consecuente
(while(cond,A),s)
\xrightarrow{\tau}
(A ; (while(cond,A),s))
}{5}$

\\

\hspace{1cm}\textbf{(While2)}
$\reglaa{ % Antecedente
\neg cond(s)
}
{ % Consecuente
(while(cond,A),s)
\xrightarrow{\tau}
(empty,s)
}{4}$

\\

\hspace{1cm}\textbf{(Pick)}
$(pick(\{(pl_i,op_i,v_i,A_i)\}_{i=1}^n ,A,t),s)
\xrightarrow{pick(pl_i,op_i,v'_i,A_i)}
(A_i,s')$

\\
\hspace{1cm}$\textrm{where}\ {\it t \geq 1, v_i \in Var, v'_i \in \entero, op_i \in OPS, pl_i \in PL,}\ \textrm{and} \ {\it s'=(op_i(\sigma[v'_i/v_i]),\rho)}.$
\\

\hspace{1cm}\textbf{(CR)}
$(createResource(EPR,val,t,A_{e{_i}}),s)
\xrightarrow{createResource(EPR,val,t,A_{e{_i}})}
(empty,s')
$\\
\hspace{0.8cm}$\textrm{where}\ {\it t\geq 1, val \in \entero}\ \textrm{and}\ \it{s'=(\sigma, \rho \cup \{EPR,val,\emptyset,t,A_{e{_i}}\})},\ \textrm{if}\ {\it EPR \notin \rho}.\
\textrm{Otherwise, } {\it \rho'=\rho}.$

\\

\hspace{1cm}\textbf{(GetProp)}
$\reglaa{ % Antecedente
{\it s=(\sigma,\rho), EPR \in \rho}      
}
{ % Consecuente
(getProp(EPR,v),s)
\xrightarrow{getProp(EPR,v')}
(empty,s')
}{6}$
\\
\hspace{1cm}where  ${\it v \in Var, v' \in \entero}$ and ${\it s'=(\sigma[v'/v],\rho)}$.
\\

\hspace{1cm}\textbf{(GetProp2)}
$\reglaa{ % Antecedente
{\it s=(\sigma,\rho), EPR \notin \rho}      
}
{ % Consecuente
(getProp(EPR,v),s)
\xrightarrow{throw}
(empty,s)
}{5}$

\\

\hspace{1cm}\textbf{(SetTime)}
$\reglaa{ % Antecedente
{\it s=(\sigma,\rho), EPR \in \rho}          
}
{ % Consecuente
(setTimeout(EPR,t), s)
\xrightarrow{setTimeout(EPR,t)}
(empty,s')
}{6.5}$\\
\hspace{1cm}where  ${\it t\geq1}$,$\ {\it s'=(\sigma,\rho[t/EPR]_2)}$.

\\

\hspace{1cm}\textbf{(SetTime2)}
$\reglaa{ % Antecedente
{\it s=(\sigma,\rho), EPR \notin \rho}          
}
{ % Consecuente
(setTimeout(EPR,t), s)
\xrightarrow{throw}
(empty,s)
}{5}$

\\

\hspace{1cm}\textbf{(SetTime3)}
$\reglaa{ % Antecedente
{\it s=(\sigma,\rho), EPR \in \rho}          
}
{ % Consecuente
(setTimeout(EPR,0), s)
\xrightarrow{throw}
(empty,s)
}{6}$

\\

\end{tabular}
}
}
\caption{\label{tran1}Action and delay transition rules without notifications.}
\end{table}

When a resource has used up its lifetime or when a subscription condition holds, a specific notification is sent to the corresponding resource subscribers, which is captured by the rules in Table \ref{tab:noti}. In these rules, the parallel operator has been extended to spawn some event handling activities, which must run in parallel with the normal activity of an orchestrator. 
% but if $(A || (A_{ej1} ( \ldots || A_{ejm})))$ then, those event handling activities that were already in $A$ will not be spawned twice. 
We therefore introduce the rules by using the following syntax for the activities in execution: $\it{(A,\mathcal{A}_e)}$, where ${\it A}$ represents the normal system workflow, and ${\it \mathcal{A}_e=\{A_{e_{i}}\}_{i=0}^m}$ are the handling activities in execution. Given any activity ${\it A}$, we write for short ${\it A||\mathcal{A}_e}$ to denote ${\it (A || (A_{e{_1}} ||( \ldots || A_{e{_m}})))}$. We assume in this operator that those event handling activities that were already in $\mathcal{A}_e$ will not be spawned twice. 
%Esto ya se metera donde se necesite
%Initially, s=($A, \emptyset$), where $A$ is the initial activity of the orchestration and the set of event handling activities is empty. 

\begin{definition}[Operational semantics with notifications]\begin{rm}
We extend both types of transition to act on pairs ${\it (A,\mathcal{A}_e)}$. The transitions have now the following form:

%\begin{figure}
%  \flushleft
%    \includegraphics[height=19cm ,width=13cm]{Figures/tabla1.eps}
%  \caption{Roles and tasks in a conference contribution management.}
%  \label{Inter}
%\end{figure}

\begin{enumerate}
\item \hspace{0.1cm}$(O:(A,\mathcal{A}_e),s)\xrightarrow{a}(O:(A',\mathcal{A}'_e),s')$, a $\in$ Act %(Action transitions).
%\item \hspace{0.1cm}$(O:(A,A_e),s)\xrightarrow{a}(O:(A,A'_e),s')$, a $\in$ Act  %\hspace{0.3cm}(Action transitions).
\item \hspace{0.1cm}$(O:(A,\mathcal{A}_e),s)\xrightarrow{}_1(O:(A',\mathcal{A}'_e),s^+)$ %\hspace{1.5cm}(Delay transitions).
\end{enumerate}
\vspace{-0.3cm}
\edfn

\begin{table}[!ht]
{\scriptsize
\framebox[12.3cm]{
\begin{tabular}{l}

\hspace{1cm}\textbf{(Throw)}
${ % Consecuente
(throw,s)
\xrightarrow{throw}
(empty, s)
}$

\hspace{0.5cm}
\textbf{(Exit)} 
$ % Consecuente
(exit,s)
\xrightarrow{exit}
(empty,s)
$
\\

\hspace{1cm}\textbf{(Receive)}
${ % Consecuente
(receive(pl,op,v),s)
\xrightarrow{receive(pl,op,v')}
(empty, s')
}$\\
\hspace{1cm}where  ${\it v \in Var, v'\in \entero, op \in OPS, pl \in PL}$, and ${\it s'=(op(\sigma[v'/v]),\rho)}$.

\\%[0.2cm]

\hspace{0.9cm}
\textbf{(Invoke)}
$
{ % Consecuente
(invoke(pl,op,v_1),s)
\xrightarrow{invoke(pl,op,v_1)}
(empty,s)
}$
%where ${\it v_1, v_2 \in Var, op \in OPS}$, and $\it{pl \in PL}$.
\\%[0.2cm]

\hspace{1cm}${\bf (\overline{Reply})}$
$
{ % Consecuente
(\overline{reply}(pl,v_2),s)
\xrightarrow{\overline{reply}(pl,v'_2)}
(empty,s')
}$\\
\hspace{1cm}where  ${\it v_2 \in Var, v'_2 \in \entero, pl \in PL}$, and ${\it s'=(\sigma[v'_2/v_2],\rho)}$. 
\\

\hspace{1cm}\textbf{(Reply)}
$
{ % Consecuente
(reply(pl,v), s)
\xrightarrow{reply(pl,v)}
(empty,s)
}$
\\

\hspace{1cm}\textbf{(Assign)}
$
{ % Consecuente
(assign(expr,v_1),s)
\xrightarrow{assign(expr,v_1)}
(empty,s')
}$\\
\hspace{1cm}where ${\it v_1 \in Var, expr}$ is an arithmetic expression, and ${\it s'=(\sigma[expr/v_1],\rho)}$.
\\

\hspace{1cm}\textbf{(Seq1)}
$\reglaa{ % Antecedente
(A_1,s)\xrightarrow{a}(A'_1,s'), a\neq exit, a\neq throw         
}
{ % Consecuente
(A_1;A_2,s)
\xrightarrow{a}
(A'_1;A_2,s')
}{5}$

\\

\hspace{1cm}\textbf{(Seq2)}
$\reglaa{ % Antecedente
(A_1,s)\xrightarrow{a}(empty,s'), a\neq exit,  a\neq throw 
}
{ % Consecuente
(A_1;A_2,s)
\xrightarrow{a}
(A_2,s')
}{5}$

%where ${\it s'=(\sigma',\rho)}$ or ${\it s'=(\sigma,\rho')}$.\\
\\

\hspace{1cm}\textbf{(Seq3)}
$\reglaa{ % Antecedente
(A_1,s)\xrightarrow{a}(empty,s), (a=throw \vee a=exit)    
}
{ % Consecuente
(A_1;A_2,s)
\xrightarrow{a}
(empty,s)
}{5}$
\\

\hspace{1cm}\textbf{(Par1)}
$\reglaa{ % Antecedente
(A_1,s)\xrightarrow{a}(A'_1,s'), a\neq exit, a\neq throw             
}
{ % Consecuente
(A_1||A_2,s)
\xrightarrow{a}
(A'_1||A_2,s')
}{5}$

%where ${\it s'=(\sigma',\rho)}$ or ${\it s'=(\sigma,\rho')}$.\\

\\

\hspace{1cm}\textbf{(Par2)}
$\reglaa{ % Antecedente
(A_2,s)\xrightarrow{a}(A'_2,s'), a\neq exit, a\neq throw            
}
{ % Consecuente
(A_1||A_2,s)
\xrightarrow{a}
(A_1||A'_2,s')
}{5}$

%where ${\it s'=(\sigma',\rho)}$ or ${\it s'=(\sigma,\rho')}$.\\

\\

\hspace{1cm}\textbf{(Par3)}
$\reglaa{ % Antecedente
(A_1,s)\xrightarrow{a}(empty,s), (a=throw \vee a=exit)            
}
{ % Consecuente
(A_1||A_2,s)
\xrightarrow{a}
(empty,s)
}{5}$

\\

\hspace{1cm}\textbf{(Par4)}
$\reglaa{ % Antecedente
(A_2,s)\xrightarrow{a}(empty,s), (a=throw \vee a=exit)                
}
{ % Consecuente
(A_1||A_2,s)
\xrightarrow{a}
(empty,s)
}{5}$

\\
\hspace{1cm}\textbf{(Par5)}
${ % Consecuente
(empty||empty,s)
\xrightarrow{\tau}
(empty,s)
}$
\\

\hspace{1cm}\textbf{(While1)}
$\reglaa{ % Antecedente
cond(s)  
}
{ % Consecuente
(while(cond,A),s)
\xrightarrow{\tau}
(A ; (while(cond,A),s))
}{5}$

\\

\hspace{1cm}\textbf{(While2)}
$\reglaa{ % Antecedente
\neg cond(s)
}
{ % Consecuente
(while(cond,A),s)
\xrightarrow{\tau}
(empty,s)
}{4}$

\\

\hspace{1cm}\textbf{(Pick)}
$(pick(\{(pl_i,op_i,v_i,A_i)\}_{i=1}^n ,A,t),s)
\xrightarrow{pick(pl_i,op_i,v'_i,A_i)}
(A_i,s')$

\\
\hspace{1cm}$\textrm{where}\ {\it t \geq 1, v_i \in Var, v'_i \in \entero, op_i \in OPS, pl_i \in PL,}\ \textrm{and} \ {\it s'=(op_i(\sigma[v'_i/v_i]),\rho)}.$
\\

\hspace{1cm}\textbf{(CR)}
$(createResource(EPR,val,t,A_{e{_i}}),s)
\xrightarrow{createResource(EPR,val,t,A_{e{_i}})}
(empty,s')
$\\
\hspace{0.8cm}$\textrm{where}\ {\it t\geq 1, val \in \entero}\ \textrm{and}\ \it{s'=(\sigma, \rho \cup \{EPR,val,\emptyset,t,A_{e{_i}}\})},\ \textrm{if}\ {\it EPR \notin \rho}.\
\textrm{Otherwise, } {\it \rho'=\rho}.$

\\

\hspace{1cm}\textbf{(GetProp)}
$\reglaa{ % Antecedente
{\it s=(\sigma,\rho), EPR \in \rho}      
}
{ % Consecuente
(getProp(EPR,v),s)
\xrightarrow{getProp(EPR,v')}
(empty,s')
}{6}$
\\
\hspace{1cm}where  ${\it v \in Var, v' \in \entero}$ and ${\it s'=(\sigma[v'/v],\rho)}$.
\\

\hspace{1cm}\textbf{(GetProp2)}
$\reglaa{ % Antecedente
{\it s=(\sigma,\rho), EPR \notin \rho}      
}
{ % Consecuente
(getProp(EPR,v),s)
\xrightarrow{throw}
(empty,s)
}{5}$

\\

\hspace{1cm}\textbf{(SetTime)}
$\reglaa{ % Antecedente
{\it s=(\sigma,\rho), EPR \in \rho}          
}
{ % Consecuente
(setTimeout(EPR,t), s)
\xrightarrow{setTimeout(EPR,t)}
(empty,s')
}{6.5}$\\
\hspace{1cm}where  ${\it t\geq1}$,$\ {\it s'=(\sigma,\rho[t/EPR]_2)}$.

\\

\hspace{1cm}\textbf{(SetTime2)}
$\reglaa{ % Antecedente
{\it s=(\sigma,\rho), EPR \notin \rho}          
}
{ % Consecuente
(setTimeout(EPR,t), s)
\xrightarrow{throw}
(empty,s)
}{5}$

\\

\hspace{1cm}\textbf{(SetTime3)}
$\reglaa{ % Antecedente
{\it s=(\sigma,\rho), EPR \in \rho}          
}
{ % Consecuente
(setTimeout(EPR,0), s)
\xrightarrow{throw}
(empty,s)
}{6}$

\\

\end{tabular}
}
}
\caption{\label{tran1}Action and delay transition rules without notifications.}
\end{table}
 
%\footnote[1]{The activities without delay transition are considered atomic.}

%\marginsize{10cm}{20cm}{10cm}{10cm}
%\newpage
\begin{table}[!ht]
{\scriptsize
\framebox[13cm]{
\begin{tabular}{ll}
\hspace{1cm}\textbf{(Subs)}
$\reglaa{ % Antecedente
{\it s=(\sigma,\rho), EPR \in \rho}          
}
{ % Consecuente
(subscribe(O,EPR,cond',A_{e{_i}}),s)
\xrightarrow{subscribe(O,EPR,cond',A_{e{_i}})}
(empty,s')
}{8}$

\\

\hspace{1cm}where ${\it s'=(\sigma,Add\_subs(\rho,EPR,O,cond',A_{e{_i}}))}$

\\[0.3cm]

\hspace{1cm}\textbf{(Subs2)}
$\reglaa{ % Antecedente
{\it s=(\sigma,\rho), EPR \notin \rho}          
}
{ % Consecuente
(subscribe(O,EPR,cond'),s)
\xrightarrow{throw}
(empty,s)
}{8}$
\\[0.3cm]

\hspace{1cm}\textbf{(Wait1D)} 
$\reglaa{ % Antecedente
t>1          
}
{ % Consecuente
(wait(t),s)
\xrightarrow{}_1
(wait(t-1), s^{+})
}{4}$

\\[0.5cm]

\hspace{1cm}\textbf{(Wait2D)} 
$
{ % Consecuente
(wait(1),s)
\xrightarrow{}_1
(empty, s^{+})
}$

\\[0.3cm]

\hspace{1cm}\textbf{(ReceiveD)}
${
(receive(pl,op,v),s)
\xrightarrow{}_1
(receive(pl,op,v), s^{+})
}$

\\[0.3cm]

\hspace{1cm}\textbf{(InvokeD)}
$
{ % Consecuente
(invoke(pl,op,v_1,v_2),s)
\xrightarrow{}_1
(invoke(pl,op,v_1,v_2), s^{+})
}$
%}

\\[0.3cm]

\hspace{1cm}\textbf{(EmptyD)} 
$
{ % Consecuente
(empty,s)
\xrightarrow{}_1
(empty, s^{+})
}$

\\[0.3cm]

\hspace{1cm}\textbf{(SequenceD)}
$\reglaa{ % Antecedente
(A_1,s)\xrightarrow{}_1(A'_1,s^{+})
}
{ % Consecuente
(A_1;A_2,s)
\xrightarrow{}_1
(A_1';A_2,s^{+})
}{4}$

\\[0.5cm]

\hspace{1cm}\textbf{(ParallelD)}
$\reglaa{ % Antecedente
(A_1,s)\xrightarrow{}_1(A'_1,s^{+}) \wedge  (A_2,s)\xrightarrow{}_1(A'_2,s^{+})            
}
{ % Consecuente
(A_1||A_2,s)
\xrightarrow{}_1
(A'_1||A'_2,s^{+})
}{5}$

\\[0.5cm]

%\multicolumn{2}{c}{
\hspace{1cm}\textbf{(Pick1D)}
$
{ % Consecuente
(pick(\{(pl_i,op_i,v_i,A_i)\}_{i=1}^n ,A,1),s)
\xrightarrow{}_1
(A,s^{+})
}$
%}

\\[0.3cm]

\hspace{1cm}\textbf{(Pick2D)}
$\reglaa{ % Antecedente
t>1           
}
{ % Consecuente
(pick(\{(pl_i,op_i,v_i,A_i)\}_{i=1}^n ,A,t),s)
\xrightarrow{}_1
(pick(\{pl_i,op_i,v_i,A_i\}_{i=1}^n ,A,t-1),s^{+})
}{10}$

\end{tabular}

}
}
\caption{\label{tran2}Action and delay transition rules without notifications.}
\end{table}

\begin{table}[!ht]
\centering 
{\scriptsize
\framebox[10cm]{
\begin{tabular}{l}

\textbf{(Notif1)}
$\reglaa{ % Antecedente
(A,s)\xrightarrow{a}(A',s'), a\neq exit, a\neq throw         
}
{ % Consecuente
(O:(A,\mathcal{A}_e),s)
\xrightarrow{a}
(O:(A',\mathcal{A}_e || N(O,s')),s')
}{6}$

\\[0.5cm]

\textbf{(Notif2)}
$\reglaa{ % Antecedente
(A_{e{_i}},s)\xrightarrow{a}(A'_{e{_i}},s'), a\neq exit, a\neq throw         
}
{ % Consecuente
(O:(A,\mathcal{A}_e),s)
\xrightarrow{a}
(O:(A,\mathcal{A}'_e || N(O,s')),s')
}{6}$

\\[0.4cm]
where $\mathcal{A}'_e=\{A'_{e{_i}}\},\ A'_{e{_i}}=A'_{e{_j}},j\neq i$. 
\\[0.2cm]

\textbf{(Notif3)}
$\reglaa{ % Antecedente
(A,s)\xrightarrow{throw}(empty,s)        
}
{ % Consecuente
(O:(A,\mathcal{A}_e),s)
\xrightarrow{throw}
(O:(A_f,\mathcal{A}_e)),s)
}{6}$

\\[0.5cm]

\textbf{(Notif4)}
$\reglaa{ % Antecedente
(A,s)\xrightarrow{exit}(empty,s)        
}
{ % Consecuente
(O:(A,\mathcal{A}_e),s)
\xrightarrow{exit}
(O:(empty,empty)),s)
}{6}$

\\[0.5cm]

\textbf{(NotifD)}
$\reglaa{ % Antecedente
(A,s)\xrightarrow{}_1(A',s^{+}), (A_{e{_i}},s)\xrightarrow{}_1(A'_{e{_i}},s^{+}), \forall i        
}
{ % Consecuente
(O:(A,\mathcal{A}_e),s)
\xrightarrow{}_1
(O:(A',\mathcal{A}'_e || T(O,s)),s^{+})
}{6}$
\\

\end{tabular}
}}
\caption{\label{tab:noti}Action and delay transition rules with notifications.}
\end{table}

Finally, the outermost semantic level corresponds to the choreographic level, which is defined upon the two previously levels. In Table \ref{tab:coreo}, we define the transition rules related to the evolution of the choreography as a whole.

%As commented when we defined the types of transitions, our operational semantics evolves at three levels, the internal one corresponds to the evolution of one activity without notifications. In the second one, we state the transitions which yield the evolution of the system in the case that some notifications are triggered, whereas the third level corresponds to the composition of different orchestrators and resources. The former two parts have been covered with the operational semantics described above. Hereby, we are going to explain in the following lines how the system evolves as a whole. In the following we will denote choreography as a composition of orchestrations and their associated web services resources.
%

\begin{definition}[Choreography operational semantics]\begin{rm}
A choreography is defined as a set of orchestrators that run in parallel exchanging messages: $C=\{O_i\}_{i=1}^c$, where $c$ is the number of orchestrators
presented in the choreography. A {\it choreography state} is then defined as follows:
$S_c=\{(O_i:(A_i,{\mathcal{A}_e}^i), s_i)\}_{i=1}^c$, where $A_i$ is the activity being performed by $O_i$ at this state, ${A_e}^i$ are the event handling activities that are currently being performed by $O_i$, and $s_i$ its current state.
\edfn

\begin{table*}[!ht]
\centering 
{\scriptsize
\framebox[12.4cm]{
\begin{tabular}{l}

\hspace{0.6cm}\textbf{(Chor1)}
$\reglaa{ % Antecedente
(O_i:(A_i,{\mathcal{A}_e}^i),s_i) \xrightarrow{exit} (O_i:(empty,empty),s_i)         
}
{ % Consecuente
\{(O_j:(A_j,{\mathcal{A}_e}^j),s_j)\}_{j=1}^c
\xrightarrow{exit}
\{(O_j:(empty,empty),s_j)\}_{j=1}^c
}{9}$\\

\\[0.3cm]

\hspace{0.6cm}\textbf{(Chor2)}
$\reglaa{ % Antecedente
(O_i:(A_i,{\mathcal{A}_e}^i),s_i) \xrightarrow{a} (O_i:(A'_i,{\mathcal{A}'_e}^i),s'_i),\ a\neq exit,\ a\neq receive,\ a\neq invoke,\ \\ \hspace{4.9cm} a\neq reply,\ a\neq \overline{reply},\ a\neq pick         
}
{ % Consecuente
\{(O_j:(A_j,{\mathcal{A}_e}^j),s_j)\}_{j=1}^c
\xrightarrow{a}
\{(O_j:(A'_j,{\mathcal{A}''_e}^j),s'_j)\}_{j=1}^c
}{10}$\\

\hspace{0.5cm}such that $A'_j=A_j,\ {\mathcal{A}''_e}^j={\mathcal{A}_e}^j||N(O_j,s'_j), \forall j\neq i, j\in\{1,\ldots,c\}$.

\\[0.4cm]

\hspace{0.6cm}\textbf{(Chor3)}
$\reglaa{ % Antecedente
(O_i:(A_i,{\mathcal{A}_e}^i),s_i) \xrightarrow{}_1 (O_i:(A'_i,{\mathcal{A}'_e}^i),{s_i}^{+}),\ \forall i \in \{1\ldots c\}, \textrm{and rules chor4,\ chor5,}\\ \hspace{7cm}   
\textrm{chor6 are not applicable}    
}
{ % Consecuente
\{(O_i:(A_i,{\mathcal{A}_e}^i),s_i)\}_{i=1}^c
\xrightarrow{}_1
\{(O_i:(A'_i,{\mathcal{A}''_e}^i),{s_i}^+)\}_{i=1}^c
}{10}$\\

\hspace{0.5cm}such that $A'_i=A_i,\ {\mathcal{A}''_e}^i={\mathcal{A}_e}^i||T(O_i,{s_i}^+)$.

\\[0.5cm]

\hspace{0.6cm}\textbf{(Chor4)}
$\reglaa{ % Antecedente
(O_i:(A_i,{\mathcal{A}_e}^i),s_i) \xrightarrow{invoke(pl,op,v_1)} (O_i:(A'_i,{\mathcal{A}'_e}^i),s_i),\ pl=(O_i,O_j),\ s_i=(\sigma_i,\rho_i),\\
(O_j:(A_j,{\mathcal{A}_e}^j),s_j) \xrightarrow{receive(pl,op,\sigma_i(v_1))} (O_j:(A'_j,{\mathcal{A}'_e}^j),s'_j)         
}
{ % Consecuente
\{(O_k:(A_k,{\mathcal{A}_e}^k)),s_k)\}_{k=1}^c
\xrightarrow{invoke(pl,op,v_1)}
\{(O_k:(A'_k,{\mathcal{A}''_e}^k),s'_k)\}_{k=1}^c
}{10}$
\\
\hspace{0.5cm}where $A'_k=A_k,\ {\mathcal{A}''_e}^k={\mathcal{A}_e}^k||N(O_k,s'_k) $ \it{if} $k\neq i,k\neq j$. 

\\[0.4cm]

\hspace{0.6cm}\textbf{(Chor5)}
$\reglaa{ % Antecedente
(O_i:(A_i,{\mathcal{A}_e}^i),s_i) \xrightarrow{reply(pl,v)} (O_i:(A'_i,{\mathcal{A}'_e}^i),s_i),\ pl=(O_i,O_j),\ s_i=(\sigma_i,\rho_i),\\
(O_j:(A_j,{\mathcal{A}_e}^j),s_j) \xrightarrow{\overline{reply}(pl,\sigma_i(v))} (O_j:(A'_j,{\mathcal{A}'_e}^j),s'_j)         
}
{ % Consecuente
\{(O_k:(A_k,{\mathcal{A}_e}^k),s_k)\}_{k=1}^c
\xrightarrow{reply(pl,\sigma_i(v))}
\{(O_k:(A'_k,{\mathcal{A}''_e}^k),s'_k)\}_{k=1}^c
}{10}$\\

\hspace{0.5cm}where $A'_k=A_k,\ {\mathcal{A}''_e}^k={\mathcal{A}_e}^k||N(O_k,s'_k) $ \it{if} $k\neq i,k\neq j$.

\\[0.4cm]

\hspace{0.6cm}\textbf{(Chor6)}
$\reglaa{ % Antecedente
(O_i:(A_i,{\mathcal{A}_e}^i),s_i) \xrightarrow{invoke(pl,op,v_1)} (O_i:(A'_i,{\mathcal{A}'_e}^i),s_i),\ pl=(O_i,O_j),\ s_i=(\sigma_i,\rho_i),\\
(O_j:(A_j,{\mathcal{A}_e}^j),s_j) \xrightarrow{pick(pl,op,\sigma_i(v_1),A)} (O_j:(A'_j,{\mathcal{A}'_e}^j),s'_j)         
}
{ % Consecuente
\{(O_k:(A_k,{\mathcal{A}_e}^k),s_k)\}_{k=1}^c
\xrightarrow{invoke(pl,op,v_1)}
\{(O_k:(A'_k,{\mathcal{A}''_e}^k),s'_k)\}_{k=1}^c
}{10}$
\\
\hspace{0.5cm}where $A'_k=A_k,\ {\mathcal{A}''_e}^k={\mathcal{A}_e}^k||N(O_k,s'_k) $ \it{if} $k\neq i,k\neq j$.
\\[0.4cm]

\end{tabular}
}}
\caption{\label{tab:coreo}Choreography transition rules.}
\end{table*}

\begin{definition}[Labeled transition system]\begin{rm}

For a choreography $C$, we define the
semantics of $C$ as the labeled transition system
obtained by the application of rules in Table \ref{tab:coreo}, starting at the state ${s_0}_c$:
\[
  {\it lts}(C) = (\mathcal{Q}, {s_0}_c,  \rightarrow)
\]

\newpage
\noindent where $\mathcal{Q}$ is the set of reachable choreography states, and
$\rightarrow \,\,= \,\,\rightarrow_1 \,\cup\, \{ \flecha{a}\,|\,$
for all basic activity $a$, or $a=\tau \,\}$.
\edfn

\begin{example}\begin{rm}Let us consider the choreography ${\it C=(O_1,O_2)}$, where \\
${\it O_i=(PL_i,Var_i,A_i,A_{f{_i}},\mathcal{A}_{e{_i}})}$, i=1, 2,${\it Var_1=\{v_1,v_3\}}$, ${\it Var_2=\{v_2,v_4\}}$, ${\it A_{f{_1}}=exit}$, and ${\it A_{f{_2}}=exit}$. Suppose that ${\it s_{0{_1}}}$ and ${\it s_{0{_2}}}$ are the initial states of $O_1$ and $O_2$, respectively, and all the variables are initially $0$. Then, ${\it A_1= assign(5,v_1);}$\\
${\it receive(pl_1,add,v_3);reply(pl_1,v_3)}$ and ${\it A_2=assign(1,v2);invoke(pl_1,add,v_2)}$. In Fig. \ref{ltsC1} we show a piece of the labeled transition
system of $C$, where:

{\small
\[
\begin{array}{ll}
A'_1 = &
{\it receive}(pl_1,add,v_3);{\it reply}(pl_1,v_3).

\\
A'_2 = &  
{\it invoke}(pl_1,add,v_2).
\\
A''_2 = &
 {\it \overline{reply}}(pl_{1},v_4).

\\
A''_1 = &  
 {\it reply}(pl_{1},v_3).

\end{array}
\]}
%and
%\[\mu_{ijk}(v_1)=i,\quad \mu_{ijk}(v_2)=j,\quad
%\mu_{ijk}(v_3)=k,\quad\mu_{ijk}(v_n)=\epsilon\quad \forall v_n \neq
%{\it clock},v_1,v_2,v_3
%\]

\begin{figure}[!h]
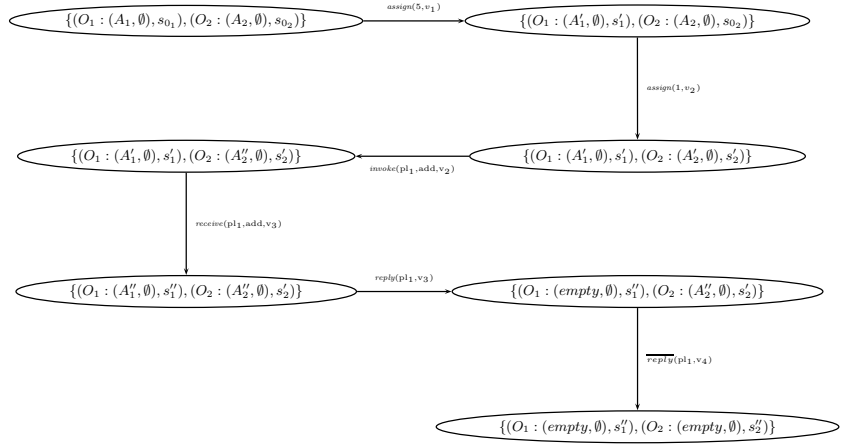

\vspace*{0.5cm}

% \hspace*{1cm}
\scalebox{0.6}{
 \pspicture(-5,7)%(10,7)
 %\psgrid[subgriddiv=1](0,0)(0,0)(15,16)
 \psset{arrows=->,linearc=.7}
 \rput(0,7){\ovalnode{N1}{$\{(O_1:(A_1,\emptyset),s_{0{_1}}), (O_2:(A_2,\emptyset),s_{0{_2}})\}$}}
 \rput(10,7){\ovalnode{N2}{$\{(O_1:(A'_1,\emptyset),s'_1), (O_2:(A_2,\emptyset),s_{0{_2}})\}$}}
 %\rput(14,7){\ovalnode{N3}{$([D_2]_\equiv,A_2,\mu_{1\epsilon\epsilon})$}}
 %\rput(1,3){\ovalnode{N4}{$([\underline{A_1}]_\equiv,A_2,\mu_{11\epsilon})$}}
 %\rput(7,3){\ovalnode{N5}{$([\underline{A_1}]_\equiv,A_2,\mu_{1\epsilon1})$}}
 \rput(10,4){\ovalnode{N6}{$\{(O_1:(A'_1,\emptyset),s'_1), (O_2:(A'_2,\emptyset),s'_2)\}$}}
 \rput(0,4){\ovalnode{N7}{$\{(O_1:(A'_1,\emptyset),{s'_1}), (O_2:(A''_2,\emptyset),{s'_2})\}$}}
 %\rput(6,3){\ovalnode{N8}{$\{(O_1:(A'_1,\emptyset),s'_1), (O_2:(A''_2,\emptyset),s''_2)\}$}}
 %\rput(1,-1){\ovalnode{N9}{$([\underline{A_1}]_\equiv,A_2,\mu_{11\epsilon})$}}
 \rput(0,1){\ovalnode{N10}{$\{(O_1:(A''_1,\emptyset),{s''_1}), (O_2:(A''_2,\emptyset),{s'_2})\}$}}
 %\rput(6,-4){\ovalnode{N11}{$\{(O_1:(A''_1,\emptyset),s''_1), (O_2:(A''_2,\emptyset),s''_2)\}$}}
\rput(10,-2){\ovalnode{N12}{$\{(O_1:(empty,\emptyset),s''_1), (O_2:(empty,\emptyset),s''_2)\}$}}
 \rput(10,1){\ovalnode{N13}{$\{(O_1:(empty,\emptyset),s''_1), (O_2:(A''_2,\emptyset),s'_2)\}$}}
%\rput(1,3){\ovalnode{N14}{$\{(O_1:(A'''_1,\emptyset),s'''_1), (O_2:(A'''_2,\emptyset),s'''_2)\}$}}

 \ncline{N1}{N2}
 \aput{:U}{\tiny{{\it assign}($5$,$v_1$)}}
 \ncline{N2}{N6}
 \aput{:L}{\tiny{{\it assign}($1$,$v_2$)}}
%
% \ncline{N3}{N4}
% \aput{:D}{\tiny{{\it inter}($r_1$,$r_2$,$v_1$,$v_2$,1)}}
%
% \ncline{N3}{N6}
% \aput{:D}{\tiny{{\it inter}($r_1$,$r_3$,$v_1$,$v_3$,2)}}
 %
% \ncline{N3}{N6}
% \aput{:U}{\tiny{{\it assign}($1$,$v_2$)}}

% \ncline{N3}{N6}
% \aput{:U}{\tiny{$\begin{array}{l}
% \quad \\
% \quad\\
% \quad \\
%  \quad \\
% \quad\\
% \quad \\ \quad \\
% +1\end{array}$}}
 %{\tiny{+1}}
%
 %\ncline{N6}{N7}
  %\aput{:L}{\tiny{$\begin{array}{l}
 %\quad \\
 %\quad\\
 %\quad \\
 % \quad \\
 %\quad\\
 %\quad \\
 % \quad \\
 %+1\end{array}$}}
 %{\tiny{+1}}
 %
 \ncline{N6}{N7}
 \aput{:D}{\tiny{{\it invoke}(pl$_{1}$,add,v$_2$)}}
  %
 %\ncline{N6}{N9}
% \aput{:D}{\tiny{{\it inter}($r_1$,$r_2$,$v_1$,$v_2$,1)}}
%
 \ncline{N7}{N10}
 \aput{:L}{\tiny{{\it receive}(pl$_{1}$,add,v$_3$)}}
%
 %\ncline{N7}{N10}
  %\aput{:L}{\tiny{$\begin{array}{l}
 %\quad \\
 %\quad\\
 %\quad \\
 % \quad \\
 %\quad\\
 %\quad \\ \quad \\
 %+1\end{array}$}}
 %{\tiny{+1}}
%
\ncline{N10}{N13}
 \aput{:U}{\tiny{{\it reply}(pl$_{1}$,v$_3$)}}
\ncline{N13}{N12}
 \aput{:L}{\tiny{{\it $\overline{reply}$}(pl$_1$,v$_4$)}}

\endpspicture}

 \vspace{1cm}

\caption{A piece of ${\it lts}(C)$ without notifications. } \label{ltsC1}
\end{figure}
\eex

\section{Case study: Online auction service}\label{cs}
The case study concerns a typical online auction process, which consists of three participants:
the online auction system and two buyers, A$_1$ and A$_2$. A seller owes a good that wants to sell to the highest possible price. Therefore, he introduces the product
in an auction system for a certain time. Then, buyers (or bidders) may place bids for the product and, when time runs out, the highest bid wins. In our case, we suppose the resource is the product for auction, the value of the resource property is the current price (only the auction system can modify it), the resource subscribers will be the buyers, their subscription conditions hold when the current product value is higher than their bid, and the resource lifetime will be the time in which the auction is active. Finally, when the lifetime has expired, the auction system sends a notification to the buyers with the result of the process (the identifier of the winner, $v_w$) and, after that, all the processes finish. Let us consider the choreography ${\it C=(O_{sys},O_1,O_2)}$, where 
${\it O_i=(PL_i,Var_i,A_i,A_{f{_i}},\mathcal{A}_{e{_i}})}$,~i=1,2,~${\it Var_{sys}=\{v_w,v_{EPR},}$
${\it end\_bid\}}$,\\
~${\it Var_1=\{v_1,v_{w{_1}}\},\ Var_2=\{v_2,v_{w_{2}}\},\ A_{f{_1}}=exit,}$ and
${\it A_{f{_2}}=exit}$. Variable $v_{EPR}$ serves to temporarily store the value of the resource property before sending; $v_1$, $v_2$, $v_{w_{}}$, $v_{w_{1}}$, $v_{w_{2}}$ are variables used for the interaction among participants, and, finally, $end\_bid$ is reset when the auction lifetime expires. Suppose ${\it s_{0{_{sys}}}, s_{0{_1}}}$ and ${\it s_{0{_2}}}$ are the initial states of $O_{sys}$, $O_1$ and $O_2$, respectively, and all the variables are initially $0$: \\[-0.2cm]

\noindent ${\it A_{sys}=assign(1,end\_bid);createResource(EPR,25,48,A_{not});}$${\it while(end\_bid > 0,A_{bid})}$.\\
${\it A_1=subscribe(O_1,EPR,EPR>=0,A_{cond_{1}});}$ ${\it while(v_{w{_1}}==0,A_{pick_{1}})}$\\
%${\it pick((pl_1,bid\_up,v_1,reply(pl_1,v_1+rand())),(pl_1,bid\_finish,v_w,empty),empty,timeout))}$\\ and
${\it A_2=subscribe(O_2,EPR,EPR>=0,A_{cond_{2}});}$${\it while(v_{w{_2}}==0,A_{pick_{2}})}$, being:\\
${\it A_{not}=assign(0,end\_bid); (invoke(pl_3,bid\_finish_1,v_w)||}$${\it invoke(pl_4,bid\_finish_2,v_w)})$\\
${\it A_{bid}= pick((pl_1,cmp,v_1,setProp(EPR,v_{EPR})),}$${\it (pl_2,cmp,v_2,setProp(EPR,v_{EPR})),}$\\$\indent \hspace{0.5cm}{\it empty,48)}$\\ 
${\it A_{cond_{1}}=getProp(EPR,v_{EPR});invoke(pl_1,{bid\_up}_1,v_{EPR})}$\\
${\it A_{cond_{2}}=getProp(EPR,v_{EPR});invoke(pl_2,{bid\_up}_2,v_{EPR})}$\\
${\it A_{pick_{1}}= pick((pl_1,bid\_up_1,v_1,{\it invoke}(pl_1,cmp,v_1);subscribe(O_1,EPR,EPR>=}$${\it v_1,}$\\$\indent \hspace{0.9cm}$ 
${\it ,A_{cond_{1}})),(pl_3,bid\_finish_1,v_1,empty),empty,48)}$\\
${\it A_{pick_{2}}= pick((pl_2,bid\_up_2,v_2,{\it invoke}(pl_2,cmp,v_2);subscribe(O_2,EPR,EPR>=}$${\it v_2,}$\\$\indent \hspace{0.9cm}$ 
${\it ,A_{cond_{2}})),(pl_4,bid\_finish_2,v_2,empty),empty,48)}$\\

%${\it A_{pick_{2}}= pick((pl_2,bid\_up_2,v_2,reply(pl_2,v_2)),}$\\ 
%\indent \hspace{1cm}${\it (pl_2,bid\_finish_2,v_2,empty),empty,48)}$\\
 
In Fig. \ref{auct} we show a part of the labeled transition
system of $C$, where:\\[-0.1cm]

{\small
\[
\begin{array}{ll}
A'_{sys} = 
{\it while}(end\_bid > 0,A_{bid}).
\\
A'_1 =   
{\it while}(v_{w{_1}}==0,A_{pick_{1}})
\\
A'_2 = 
{\it while}(v_{w{_2}}==0,A_{pick_{2}})
%wait(10)})
\\
A''_1 =  
{\it A_{pick_{1}};while}(v_{w{_1}}==0,A_{pick_{1}})
\\
A''_2 = 
{\it A_{pick_{2}};while}(v_{w{_2}}==0,A_{pick_{2}})

\\
A''_{sys} =   
{\it A_{bid};while}(end\_bid > 0,A_{bid}).
\end{array}
\]}
\\
Let us note that the operations $bid\_up_{1}$ and $bid\_up_{2}$ are used to increase the current bid by adding a random amount to the corresponding variable $v_i$, the operations $bid\_finish_{1}$, $bid\_finish_{2}$ reset the value of $v_w$ to finish both buyers. Finally, $cmp$ is an auction system operation that receives as parameter a variable of the buyers, $v_i$, and if the variable value is greater than the current value of $v_{EPR}$, then $v_{EPR}$ is modified with this new value. After that, by means of the activity $setProp(EPR,v_{EPR})$, we can update the value of the resource property with the new bid.

\twocolumn
\begin{landscape}
\begin{figure*}[!ht]
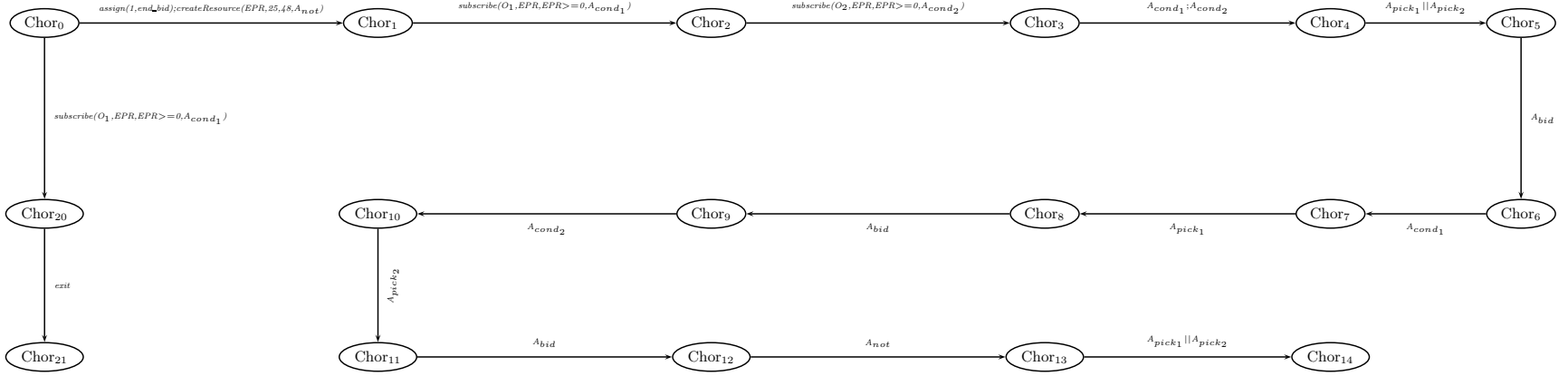

\vspace*{6cm}

 \hspace*{-1cm}
\scalebox{0.7}{
 \pspicture(1,7)%(10,7)
 %\psgrid[subgriddiv=1](0,0)(0,0)(15,16)
 \psset{arrows=->,linearc=.7}
 \rput(-12,17){\ovalnode{N1}{Chor$_0$}}
 \rput(-5,17){\ovalnode{N2}{Chor$_1$}}
 \rput(-12,10){\ovalnode{N4}{Chor$_{21}$}}
 \rput(2,17){\ovalnode{N6}{Chor$_2$}}
 \rput(9,17){\ovalnode{N7}{Chor$_3$}}
 \rput(15,17){\ovalnode{N10}{Chor$_4$}}
 \rput(19,13){\ovalnode{N12}{Chor$_6$}}
 \rput(19,17){\ovalnode{N13}{Chor$_5$}}
 \rput(-12,13){\ovalnode{N14}{Chor$_{20}$}}
 \rput(15,13){\ovalnode{N15}{Chor$_{7}$}}
 \rput(9,13){\ovalnode{N16}{Chor$_{8}$}}
 \rput(2,13){\ovalnode{N17}{Chor$_{9}$}}
 \rput(-5,13){\ovalnode{N18}{Chor$_{10}$}}
 \rput(-5,10){\ovalnode{N19}{Chor$_{11}$}}
 \rput(2,10){\ovalnode{N20}{Chor$_{12}$}}
 \rput(9,10){\ovalnode{N21}{Chor$_{13}$}}
\rput(15,10){\ovalnode{N22}{Chor$_{14}$}}
%\rput(14,-1){\ovalnode{N23}{Chor$_{15}$}}
%\rput(18,-1){\ovalnode{N24}{Chor$_{16}$}}
%\rput(22,-1){\ovalnode{N25}{Chor$_{17}$}}
%\rput(11,-1){\ovalnode{N26}{Chor$_{18}$}}
%\rput(6,-5){\ovalnode{N27}{Chor$_{19}$}}
%\rput(0,-1){\ovalnode{N28}{Chor$_{22}$}}
%\rput(0,1){\ovalnode{N29}{Chor$_{23}$}}

 \ncline{N1}{N2}
 \aput{:U}{\tiny{{\it assign(1,end\_bid);createResource(EPR,25,48,A$_{not}$)}}}
 \ncline{N2}{N6}
 \aput{:U}{\tiny{{\it subscribe(O$_1$,EPR,EPR$>=$0,A$_{cond_{1}}$)}}}

 \ncline{N14}{N4}
 \aput{:L}{\tiny{{\it exit}}}
%
% \ncline{N3}{N6}
% \aput{:D}{\tiny{{\it inter}($r_1$,$r_3$,$v_1$,$v_3$,2)}}
 %
% \ncline{N3}{N6}
% \aput{:U}{\tiny{{\it assign}($1$,$v_2$)}}

% \ncline{N3}{N6}
% \aput{:U}{\tiny{$\begin{array}{l}
% \quad \\
% \quad\\
% \quad \\
%  \quad \\
% \quad\\
% \quad \\ \quad \\
% +1\end{array}$}}
 %{\tiny{+1}}
%
 \ncline{N6}{N7}
 \aput{:U}{\tiny{{\it subscribe(O$_2$,EPR,EPR$>=$0,A$_{cond_{2}}$)}}}
 
 \ncline{N7}{N10}
 \aput{:U}{\tiny{{\it A$_{cond_{1}};$A$_{cond_{2}}$}}}

%\ncline{N7}{N10}
%\nccircle[angleA=2,nodesep=4pt,linestyle=dashed,dash=2mm 2mm 0.1 0.2,linetype=0]{->}{N7}{.6}
% \bput{:D}{\tiny{+1}}
 
\ncline{N10}{N13}
\aput{:U}{\tiny{{\it A$_{pick_{1}}||$A$_{pick_{2}}$}}}
%\aput{:U}{\tiny{{\it A_{pick_{1}} }(pl$_1$,EPR$_1)$$||${\it reply}(pl$_2$,EPR$_1$)}}

%
\ncline{N13}{N12}
\aput{:L}{\tiny{{\it A$_{bid}$}}}
%\aput{:L}{\tiny{{\it reply}(pl$_1$,v$_1$)$||${\it reply}(pl$_2$,v$_2$)}} 
%
\ncline{N1}{N14}
 \aput{:L}{\tiny{{\it subscribe(O$_1$,EPR,EPR$>=$0,A$_{cond_{1}}$)}}}

\ncline{N12}{N15}
\aput{:D}{\tiny{{\it A$_{cond_{1}}$}}}
%\aput{:D}{\tiny{{\it A$_{bid}$;A$_{bid}$}}}

\ncline{N15}{N16}
\aput{:D}{\tiny{{\it A$_{pick_{1}}$}}}

\ncline{N16}{N17}
\aput{:D}{\tiny{{\it A$_{bid}$}}}
%\aput{:D}{\tiny{{\it A$_{pick_{1}}$}}}

\ncline{N17}{N18}
\aput{:D}{\tiny{{\it A$_{cond_{2}}$}}}
%\aput{:D}{\tiny{{\it A$_{bid}$;A$_{bid}$}}}

\ncline{N18}{N19}
\aput{:D}{\tiny{{\it A$_{pick_{2}}$}}}

\ncline{N19}{N20}
\aput{:U}{\tiny{{\it A$_{bid}$}}}

\ncline{N20}{N21}
\aput{:U}{\tiny{{\it A$_{not}$}}}

\ncline{N21}{N22}
\aput{:U}{\tiny{{\it A$_{pick_{1}}||$A$_{pick_{2}}$}}}

%\ncline{N22}{N23}
%\aput{:U}{\tiny{{\it reply}(pl$_2$,v$_2$+rand())}}

%\ncline{N23}{N24}
%\aput{:D}{\tiny{{\it receive}(pl$_2$,v$_2$)}}

%\ncline{N24}{N25}
%\aput{:D}{\tiny{{\it reply}(pl$_1$,v$_w$)$||${\it reply}(pl$_2$,v$_w$)}} 

%\ncline{N25}{N26}
%\aput{:D}{\tiny{{\it pick}(pl$_1$,bid$\_$finish,v$_w$,empty) $||${\it pick}(pl$_2$,bid$\_$finish,v$_w$,empty)}}

%\ncline{N26}{N27}
%\aput{:D}{\tiny{{\it empty}$||${\it empty}}}

\endpspicture}

 \vspace{-6cm}

\caption{A piece of ${\it lts}(C)$ for the online auction service.} \label{auct}
 \vspace{1cm}

\[
\begin{array}{ll}
\hspace{-10cm}Chor_0 = \{(O_{sys}:(A_{sys},\emptyset),s_{0{_{sys}}}),(O_1:(A_1,\emptyset),s_{0{_1}}), (O_2:(A_2,\emptyset),s_{0{_2}})\}
&
\hspace{-1cm}Chor_1 = \{(O_{sys}:(A'_{sys},\emptyset),s'_{0{_{sys}}}),(O_1:(A_1,\emptyset),s_{0{_1}}), (O_2:(A_2,\emptyset),s_{0{_2}})\}
\\
\hspace{-10cm}Chor_2 = \{(O_{sys}:(A'_{sys},\emptyset),s''_{0{_{sys}}}),(O_1:(A'_1,\emptyset),s_{0{_1}}), (O_2:(A_2,\emptyset),s_{0{_2}})\}
&
\hspace{-1cm}Chor_3 = \{(O_{sys}:(A'_{sys},A_{cond_{1}};A_{cond_{2}}),s'''_{0{_{sys}}}),(O_1:(A'_1,\emptyset),s_{0{_1}}),(O_2:(A'_2,\emptyset),s_{0{_2}})\}
\\
\hspace{-10cm}Chor_4 = \{(O_{sys}:(A'_{sys},\emptyset),s'''_{0{_{sys}}}),(O_1:(A''_1,\emptyset),s_{0{_1}}), (O_2:(A''_2,\emptyset),s_{0{_2}})\}
&
\hspace{-1cm}Chor_5= \{(O_{sys}:(A''_{sys},\emptyset),s'''_{0{_{sys}}}),(O_1:(A'_1,\emptyset),s_{0{_1}}), (O_2:(A'_2,\emptyset),s_{0{_2}})\}
\\
\hspace{-10cm}Chor_6 = \{(O_{sys}:(A'_{sys},A_{cond_{1}}),s'''_{0{_{sys}}}),(O_1:(A'_1,\emptyset),s_{0{_1}}),(O_2:(A'_2,\emptyset),s_{0{_2}})\}
&
\hspace{-0.5cm}Chor_7 = \{(O_{sys}:(A'_{sys},\emptyset),s'''_{0{_{sys}}}),(O_1:(A''_1,\emptyset),s_{0{_1}}), (O_2:(A'_2,\emptyset),s_{0{_2}})\}
\\
\hspace{-10cm}Chor_8 =\{(O_{sys}:(A''_{sys},\emptyset),s'''_{0{_{sys}}}),(O_1:(A'_1,\emptyset),s_{0{_1}}),(O_2:(A'_2,\emptyset),s_{0{_2}})\}
&
\hspace{-1cm}Chor_9 =\{(O_{sys}:(A'_{sys}, A_{cond_{2}}),s'''_{0{_{sys}}}),(O_1:(A'_1,\emptyset),s_{0{_1}}),(O_2:(A'_2,\emptyset),s_{0{_2}})\}
\\
\hspace{-10cm}Chor_{10} =\{(O_{sys}:(A'_{sys},\emptyset),s'''_{0{_{sys}}}),(O_1:(A'_1,\emptyset),s_{0{_1}}), (O_2:(A''_2,\emptyset),s_{0{_2}})\}
&
\hspace{-1cm}Chor_{11} =\{(O_{sys}:(A''_{sys},\emptyset),s'''_{0{_{sys}}}),(O_1:(A'_1,\emptyset),s_{0{_1}}), (O_2:(A'_2,\emptyset),s_{0{_2}})\}
\\
\hspace{-10cm}Chor_{12} =\{(O_{sys}:(A'_{sys},A_{not}),s'''_{0{_{sys}}}),(O_1:(A'_1,\emptyset),s_{0{_1}}), (O_2:(A'_2,\emptyset),s_{0{_2}})\}
&
\hspace{-0.7cm}Chor_{13} =\{(O_{sys}:(empty,\emptyset),s'''_{0{_{sys}}}),(O_1:(A''_1,\emptyset),s_{0{_1}}), (O_2:(A''_2,\emptyset),s_{0{_2}})\}
\\
\hspace{-10cm}Chor_{14} = \{(O_{sys}:(empty,\emptyset),s'''_{0{_{sys}}}),(O_1:(empty,\emptyset),s_{0{_1}}),(O_2:(empty,\emptyset),s_{0{_2}})\}
&
\\
\end{array}
\]

\end{figure*}
\vspace{-5cm}

%\end{flushleft}
\end{landscape}

\onecolumn
\section{Conclusions and Future Work}\label{conclusions}
We have presented in this paper a formal model for the description of composite web services with resources associated, and orchestrated
by a well-know business process language (BPEL). The main contribution has therefore been the integration of WSRF, a resource management language, with BPEL, taking into 
account the main structural elements of BPEL, as its basic and structured activities, notifications, event handling and fault handling. Furthermore, special attention has been given to timed constraints, as WSRF consider that resources can only exist for a certain time (lifetime). Thus, resource leasing is considered in this work, which is a concept that has become increasingly popular in the field of distributed systems. To deal with notifications, event handling and fault handling, the operational semantics has been defined at three levels, the outermost one corresponding to the choreographic view of the composite web services.

As future work, we plan to extend the language with some additional elements of BPEL, such as termination and compensation handling. Compensation is an important topic in web services due to the possibility of faults. We are also working on a semantics based on timed colored petri nets.

\section*{Acknowledgement}
Partially supported by the Spanish Government (co-financed by FEDER funds)
with the project TIN2009-14312-C02-02 and the JCCLM regional project
PEII09-0232-7745.

\end{document}